# Neutron background signal in superheated droplet detectors of the Phase II SIMPLE dark matter search

A.C. Fernandes[1,*], A. Kling[1], M. Felizardo[1], T.A. Girard[1], A.R. Ramos[1], J.G. Marques[1], M.I. Prudêncio[1], R. Marques[1], F.P. Carvalho[1], I.L. Roche[2]

(for the SIMPLE Collaboration)

[1] Centro de Ciências e Tecnologias Nucleares (C$^2$TN), Instituto Superior Técnico, Universidade de Lisboa. Estrada Nacional 10 (km 139.7), 2695-066 Bobadela LRS – Portugal

[2] Laboratoire Souterrain à Bas Bruit, University of Nice, University of Avignon, Centre National de la Reserche Scientifique, Aix-Marseille University, Observatoire de la Côte d'Azur, 84400 Rustrel - France

* corresponding author: anafer@ctn.tecnico.ulisboa.pt

**ABSTRACT**

The simulation of the neutron background for Phase II of the SIMPLE direct dark matter search experiment is fully reported with various improvements relative to previous estimates. The model employs the Monte Carlo MCNP neutron transport code, using as input a realistic geometry description, measured radioassays and material compositions, and tabulated ($\alpha$,n) yields and spectra. Developments include the accounting of recoil energy distributions, consideration of additional reactions and materials and examination of the relevant ($\alpha$,n) data. A thorough analysis of the simulation results is performed that addresses an increased number of non-statistical uncertainties. The referred omissions are seen to provide a net increase of 13% in the previously-reported background estimates whereas the non-statistical uncertainty rises to 25%. The final estimated recoil event rate is 0.372 $\pm$ 0.002 (stat.) $\pm$ 0.097 (non-stat.) evt/kgd resulting in insignificant changes over the results of the experiment.

## 1. Introduction

Direct dark matter search experiments are based on the observation of nuclear recoils induced by the elastic scattering of Weakly Interacting Massive Particles (WIMPs) with target nuclei. SIMPLE (Superheated Instrument for Massive Particle Search) is one of three international experiments using superheated liquids: SIMPLE and PICASSO have employed superheated droplet detectors (SDDs) while COUPP uses bubble chambers for increased active mass and has recently merged with PICASSO into PICO [1-4].

A SDD is a suspension of micrometric superheated liquid droplets in a viscoelastic gel which undergo transitions to the gas phase upon absorption of energy from radiation [5]. A proto-bubble will expand to a millimeter-sized bubble when two thermodynamically-defined threshold conditions are satisfied with respect to (i) the energy deposited and (ii) the distance within the droplet where the deposition occurs [6]. SIMPLE employs $C_2ClF_5$ in a food gel-based matrix and operates at 9 $^{o}$C and 2 bar. A critical energy deposition $E_{crit}$=8.2 keV and Linear Energy Transfer $LET_{crit}$=159.7 keV $\mu m^{-1}$ are required for a bubble nucleation, which is detected via an acoustic instrumentation and signal analyses that confer and identify its characteristic frequency signature [7].

Expected WIMP-nucleon interaction rates are extremely small, below $10^{-3}$ event (evt) per kilogram of detector mass and day (kgd). The key to the direct search experiments lies in the ability to suppress and reject more prevalent signals originating in cosmic radiation and natural radioactivity. Background reduction is generally achieved through operation in underground facilities, additional radiation shielding, purification of detector components and radon purging. As the bubble nucleation conditions render SIMPLE SDDs intrinsically insensitive to low LET radiations (e.g. photons and electrons), the detector background signal is restricted to alpha particles and neutrons. Nuclear recoil and alpha events are discriminated based on signal amplitudes: alphas produce larger amplitudes as a result of a higher bubble expansion power from the formation of various proto-bubbles [1, 8, 9].

The assessment of the neutron background is crucial because fast neutrons interact with nuclei through elastic scattering to generate nuclear recoils which mimic the characteristic signal expected from WIMPs. In deep underground facilities neutrons originate primarily from emitters present at trace levels in the facility and detector materials. Among the naturally occurring emitters, $^{238}$U and $^{232}$Th decay chains are the most important, producing neutrons both by spontaneous fission (with low branching ratios) and ($\alpha$,n) reactions – alphas being emitted by the various descendants in the decay series. Neutron energies up to 10-20 MeV are involved, which are sufficiently high to cross significant shielding thicknesses.

In a direct dark matter search, the reduced neutron field intensity hampers the experimental determination of the neutron background due to, e.g., insufficient sensitivity, electronic noise or dose during transport. Simulation techniques have a leading role in the determination of the neutron background, namely through the application of general-purpose Monte Carlo (MC) neutron transport simulation codes. MCNP, GEANT4 and FLUKA [10-12] have been used by most experiments to optimize the neutron shielding, simulate detector responses, interpret the detector signal, select the experiment materials, and estimate or calculate the neutron-induced detector background signal [13-24]. The application of MCNP is restricted to the natural radioactivity, whereas FLUKA and GEANT4 offer the possibility to calculate both the muon-induced and the "local" neutron component, and can be applied to a broad scope of interactions.

The above codes can model particle transport, production of secondaries and particle detection. However, their modelling of neutron production through spontaneous fission or ($\alpha$,n) reactions is either inexistent or inaccurate at the energy range relevant for the natural background studies, and neutron yields and spectra are extracted more conveniently from evaluated data. The SOURCES code [25] is often used to derive the ($\alpha$,n) spectra and production yields due to the decay of various radionuclides. As the original version is restricted to alpha energies up to 6.5 MeV, in-house modifications are developed by some groups [26, 27] to include the higher-energy Po descendants: $^{212/216}$Po from $^{232}$Th decay and $^{214}$Po from $^{238}$U decay). In this work we followed the same approach as LUX [23] and used the application developed at the University of South Dakota (USD) [28] to obtain the ($\alpha$,n) data, assuming secular equilibrium (equal activities) among the various $^{238}$U and $^{232}$Th descendants.

MCNP simulations were initially used in an optimization study towards the design of a neutron shield [29] for SIMPLE Phase II with a simplified event rate calculation. Improved estimates of the recoil background [30] were extracted in publications of the experiment results [1, 8, 31]. Still, maximum neutron energy transfer was assumed, which vaguely compensated the missing contributions from materials and nucleation-inducing reactions that had not been characterized or evaluated. With the uncertainties in the USD ($\alpha$,n) data being unreported, those from SOURCES were assigned [32]. In this work, we provide a more complete calculation of the event rate that (i) considers the recoil energy distribution, (ii) adds the contribution of inelastic scattering reactions, (iii) includes more materials - namely, heterogeneous materials whose examination is not straightforward - and (iv) examines the quality of the USD ($\alpha$,n) yields and spectra. The uncertainty analysis is improved by the inclusion of the ($\alpha$,n) evaluation results and new features: non-equilibrium in the decay chains, simulations accuracy and a significant number of other, albeit smaller contributions. A full, detailed report of the background estimate for Phase II of SIMPLE is presented herein.

## 2. Experimental set-up

SIMPLE was run in the GESA facility of Laboratoire Souterrain à Bas Bruit (LSBB, southern France) [33]. GESA is a 60m$^3$ room at a depth of 500 m within LSBB, corresponding to 1500 m water-equivalent (mwe) depth. Floor plan dimensions are (400 x 564) cm$^2$. The ceiling has a semi-cylindrical shape (diameter 404 cm), the room height varying between 212 and 305 cm. The surrounding rock is calcite. Room walls, ceiling and floor consist of concrete, with a thickness between 30 and 100 cm. The room is equipped with a 1 cm-thick steel lining forming a Faraday cage that shields against electromagnetic noise.

Each SDD consisted of a 900 ml glycerin-based gel matrix with a 1-2 wt.% suspension of $C_2ClF_5$ droplets (average radius of 30 μm). The detector container was a (9 x 9 x 12) cm$^3$ flask of laboratory borosilicate glass (BSG), with a thickness of 5 mm. A 2 cm glycerin layer above the gel embeds the microphone employed in the acoustic instrumentation and reduces the diffusion of atmospheric radon.

The experiment used fifteen SDDs installed in a water bath with a cross-sectional dimension of (97 x 130) cm$^2$, in the central region of the room. The SDDs were distributed in alternating positions in a 16 cm square lattice. The water level was maintained 3 cm above the glycerin level further reducing radon diffusion and improving the thermal equilibrium in the detector. The tank walls consisted of glass-reinforced plastic (GRP) with a thickness of 5 mm, surrounded by 5 cm of polyurethane foam for acoustic insulation; neither was previously included in background estimates.

GESA's room floor contains several steel-covered, 50 cm deep crawl spaces previously used for cable conduits. Some of them are located around the tank, which is supported by a concrete pedestal. For a more detailed description of the disposition of Phase II see Refs. 1, 8, 29 and 31.

A preliminary evaluation of the original SIMPLE set-up in GESA (Phase I, 2004-2007) [34] demonstrated the need to implement a neutron shield for reducing the neutron-induced event rate well below 1 evt/kgd [29]. Phase II of SIMPLE (2009-2011) comprised two stages of measurements [8, 31] performed under a shielded configuration using water as neutron moderator. The neutron shielding consisted of a pile of 20 liter water boxes (22 x 25 x 38) cm$^3$, symmetrically installed around and above the tank to produce water thicknesses of 50-75 cm, respectively. The SDDs were raised 50 cm above the tank floor for additional shielding from neutrons emitted by the concrete pedestal. The pedestal is surrounded by water boxes, producing a shield of 44 cm width and 50 cm height.

Stage 1 (2009-2010) was carried out with a net exposure of 11.53 kgd achieving a measured recoil event rate of 0.61 evts/kgd. The calculation of the neutron background suggested that a relevant improvement in the shielding would be gained using additional hydrogen-based shielding to reduce the neutron contribution from the pedestal: the tank (originally sitting on a wooden support structure with 18 cm thickness) was raised above the concrete floor in order to accommodate a polyethylene block of 10 cm thickness and a wooden layer (2 cm) between the tank and the wooden support. The free spaces underneath the wooden platform were filled with additional polyethylene (10 cm thickness), paraffin (8 cm) and wood (4 cm). Finally, as the water boxes become slightly deformed due to the weight loading, direct channels across the lateral shielding were eliminated with an overlapping box arrangement. The last stage of measurements (Stage 2, 2010-2011) involved an exposure of 6.71 kgd and 0.15 evts/kgd from recoils.

Figure 1 presents the geometrical model employed in the simulations of the SIMPLE set-up in GESA. Some structures were removed from the figure for clarity (rock, concrete and steel in the front and right walls, water boxes in the front and right side of the tank and the tank water surrounding the detectors). Surrounding the cavern, uniform layers of concrete and rock were considered with thicknesses of 30 cm and 1 m respectively, after evaluating the influence of these parameters on the calculated background [29].

### 3. Simulation strategy

Version 4C of MCNP was used in the simulations. Transport cross sections were taken mostly from the ENDF-B6.0 library included in the package.

Fluence rates were calculated using the track-length estimator of MCNP (F4 tally) in the active volume of the detectors. MCNP yields group fluence rates, i.e. fluence rates of neutrons with energies within the limits of a user-defined bin structure. Our group structure consisted of equi-lethargy groups in number of three per decade between 0.1 meV and 1 keV, and ten per decade from 1 keV to 20 MeV.

The calculation of reaction rates inducing bubble nucleations in the superheated liquid inducing nucleations (Section 7.1) was separated from the MCNP simulations in order to facilitate the evaluation of aspects related to reaction cross section data. The FLXPRO code of the LSL-M2 package [35] was employed to determine the average cross sections corresponding to our group structure. Reaction cross sections and uncertainties were obtained from the ENDF-B7.1 library [36]. The energy distribution of the recoiling target nuclei, not previously included, was considered at a final stage using the full version of the SPECTER code [37].

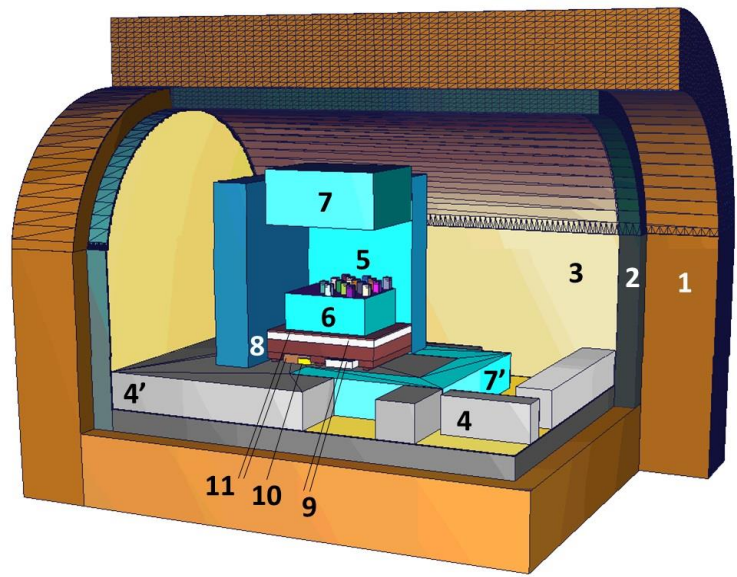

**Fig. 1.** Geometry model for the MCNP calculations of the neutron background in the final stage of SIMPLE Phase II. 1: rock; 2: concrete walls, floor and ceiling; 3: steel lining; 4: concrete floor structures defining cable conduits; 4': tank pedestal; 5: detectors; 6: tank (water below the detectors); 7: water shield around and above the tank; 7': water shield around the tank pedestal; 8: wood support; 9: polyethylene shield (layer and insert); 10: paraffin insert; 11: wood layer and insert.

MCNP outputs are given relative to one source neutron, and must be scaled to the actual source emission rate in order to obtain absolute results. Besides its emission rate, the source is characterized in terms of location, energy and angular distribution of the emitted neutrons.

Theoretical (analytical or tabulated) distributions were used to describe the energy spectrum of the source neutrons. The contributions of the $^{238}$U and $^{232}$Th decay chains in each material were considered individually, assuming uniform emitter distribution and isotropic emission. The various source materials and reactions were also discriminated in order to evaluate their relative contribution to the overall event rate.

### 4. Material data

**4.1 Composition**

Material composition plays an important role in neutron attenuation and production. Light nuclei are the most relevant, since they maximize the neutron energy transfer in elastic collisions, and may exhibit high cross sections for ($\alpha$,n) reactions. For SIMPLE it was essential to quantify (i) hydrogen in wood allowing to model the moderation of neutrons emitted by the concrete pedestal, and (ii) boron in BSG in order to define the production of ($\alpha$,n) neutrons in the detector containers.

Ion beam-based techniques are particularly adequate to quantify light elements, and were applied at C$^2$TN, Portugal. Elastic recoil detection using a 2 MeV He$^+$ beam was applied to determine the $^1$H content in wood, the remaining elements being quantified simultaneously by Rutherford Backscattering (RBS). The amount of $^{11}$B in glass was measured by Nuclear Reaction Analysis and Resonant Elastic Scattering, while the other elements were determined by RBS in the same run. For optimization of the detection conditions a 1.55 MeV proton beam was employed for these measurements.

Chemical analyses performed at the University of Avignon, France yielded the mineral composition of rock and concrete.

Standard compositions for air, water, glycerin, paraffin and polyethylene were used [38]; the composition of the steel GESA lining and the microphone were assumed to be 100% iron.

The GRP forming the tank walls was assumed to be the commonly used polypropylene reinforced with fibers of E-glass whose density and composition ranges were obtained from reference data [39]; the average values of the range limits were used in the model. The E-glass (typically 10-30 wt.% in GRP) was not included in the GRP description for the MCNP simulations. A microscope image of the GRP (Fig. 2) confirms that, similarly to other GRPs [40], the individual fibers form a compact roving which can be considered as a unique fiber. The measured diameters of the individual fiber and roving are 20$\pm$1 $\mu$m and 300$\pm$20 $\mu$m respectively.

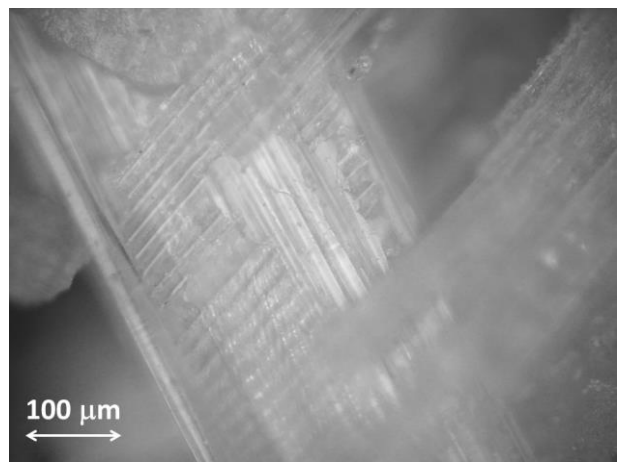

**Fig. 2**. Glass fibers emerging from the GRP, 200x. Rovings of different orientations are shown, the proximal being partly broken near its edge.

The gel matrix contains various materials of standard compositions [41] except for gelatin – its mass composition assumed to be 12.5% proline, 12.5% hydroxyproline, 20% glycine, 23% glutamic acid, 16% arginine and 16% alanine [42]. The superheated liquid was not included in the gel description for the MCNP simulations.

Standard densities were considered for steel, air, water and glycerin. The densities of wood, paraffin, polyethylene, bulk microphone, GRP, BSG, concrete and rock were measured by the immersion method, their uncertainties reflecting mainly those in the volume measurement. The density of the gel matrix was initially calculated considering the nominal mass and volume (1183 g/900 cm$^3$). The final value and its uncertainty are derived from a comparison with the density of the superheated liquid at the thermodynamic operating conditions (1.33 g cm$^{-3}$), considering that density matching is achieved for detector fabrication.

Table 1 summarizes the results. For BSG and wood a good agreement was found between measured and nominal compositions [43, 44]. The nominal data for BSG corresponds to the material used in the experiment.

**Table 1.** Materials composition and density. In the cases of borosilicate glass (BSG), E-glass and wood, nominal compositions are given in brackets. For E-glass, the range limits for $B_2O_3$ are specified.

a) Composition (wt. %) of inorganic materials

|  | Rock | Concrete | BSG |  | E-glass |
|---|---|---|---|---|---|
| $SiO_2$ |  | 37.20 | 81.8±3.3 | (81) | (55) |
| $Al_2O_3$ |  | 3.58 | 2.3±0.09 | (2) | (14) |
| $Fe_2O_3$ |  | 1.40 |  |  | (0.3) |
| MnO |  | 0.05 |  |  |  |
| MgO | 0.31 | 0.75 |  |  | (2.2) |
| CaO |  |  |  |  | (22) |
| $Ca_2CO_3$ | 99.69 | 55.42 |  |  |  |
| $Na_2O$ |  | 0.67 | 2.0±0.08 | (2) | (0.5) |
| $K_2O$ |  | 0.72 | 1.5±0.06 | (2) |  |
| $TiO_2$ |  | 0.15 |  |  | (0.5) |
| $P_2O_5$ |  | 0.06 |  |  |  |
| $B_2O_3$ |  |  | 12.4±0.50 | (13) | (5 [4-6]) |
| $F_2$ |  |  |  |  | (0.5) |

**Table 1 (cont.).**

b) Composition (wt. %) of organic materials

|   | Wood |   | Gel matrix |
|---|------|---|------------|
| H | 5.8  | (6) | 9.10 |
| C | 47.6 | (50) | 33.80 |
| O | 46.0 | (44) | 56.35 |
| K | 0.6  | (0) |  |
| N |      |   | 0.75 |

c) Material densities (g cm$^{-3}$)

| Material | Density | Material | Density |
|----------|---------|----------|---------|
| Rock | 2.61±0.01 | Gel & liquid | 1.32±0.01 |
| Concrete | 2.39±0.01 | BSG | 2.23±0.01 |
| Steel | 7.874 | GRP | 1.05±0.01 |
| Wood | 0.51±0.01 | E-glass | (2.55±0.03) |
| Water | 1.0 | Glycerin | 1.261 |
| Paraffin | 0.83±0.01 | Bulk microphone | 1.98±0.01 |
| Polyethylene | 0.95±0.01 | Air | 1 x 10$^{-3}$ |

**4.2 Radioactive contaminants**

Traces of $^{238}$U and $^{232}$Th (or their respective descendants) were quantified in most experiment materials via low detection limit techniques. The gel was evaluated in an early phase of SIMPLE by $\alpha$-spectroscopy at Pacific Northeast National Laboratory, USA. Alpha spectroscopy was also used at C$^2$TN to analyze the wood and shielding water after incineration and evaporation, respectively. The structural materials of GESA (concrete, rock and steel) were assayed by low-background gamma spectroscopy at the Laboratoire Souterrain de Modane, France. The paraffin and polyethylene used in the shielding, the GRP tank walls, the BSG detector container and the microphone were evaluated at C$^2$TN by Neutron Activation Analysis (NAA) using the comparative method, i.e., through co-irradiation with standard materials certified for $^{238}$U or $^{232}$Th concentration.

Spectroscopic methods can provide information about various radioisotopes of the $^{238}$U and $^{232}$Th decay chain and thereby allows the evaluation of secular equilibrium. In contrast, NAA can only quantify radionuclides for which certified reference materials are available.

Table 2 shows the measured amounts of $^{238}$U and $^{232}$Th in the various materials. The experimental uncertainties are reported at the 1-σ (68% confidence) level. Available information about radioactive descendants naturally produced through the decay series of $^{238}$U or $^{232}$Th is included. When the measurement is below the detection limit (e.g., paraffin, gel), the latter limit is given.

**Table 2.** Activity concentration of $^{238}$U and $^{232}$Th in GESA and SIMPLE materials (Bq l$^{-1}$ for water and Bq kg$^{-1}$ for other materials).

| Material | Method | $^{238}$U | $^{232}$Th |
|---|---|---|---|
| Concrete | gamma-spectroscopy | 10.5 ± 1.0 | 7.7 ± 0.2 |
| Gel | alpha-spectroscopy | < 9.0 x10$^{-3}$ | - |
| BSG | NAA | 2.74 ± 0.41 | 1.27 ± 0.11 |
| Microphone | NAA | < 13.6 | - |
| Paraffin | NAA | < 0.25 | < 0.4 |
| Polyethylene | NAA | 3.99x10$^{-1}$ ± 1x10$^{-3}$ | - |
| Polyurethane | NAA | 0.58 ± 0.15 | < 0.4 |
| Rock | gamma-spectroscopy | 5.0 ± 2.5 | 1.6x10$^{-1}$ ± 1.3x10$^{-2}$ |
| Steel $^{1)}$ | gamma-spectroscopy | 3.6x10$^{-2}$ ± 1.1x10$^{-2}$ | 1.3x10$^{-2}$ ± 1x10$^{-3}$ |
| GRP | NAA | 7.76 ± 0.49 | 18.0 ± 1.9 |
| Water $^{(2)}$ | alpha-spectroscopy | 3.20x10$^{-2}$ ± 6.0x10$^{-4}$ | 5.00x10$^{-5}$ ± 1.00x10$^{-5}$ |
| Wood $^{(3)}$ | alpha-spectroscopy | 1.07x10$^{-1}$ ± 1.5x10$^{-2}$ | 3.0x10$^{-3}$ ± 9.0x10$^{-4}$ |

Additional radioisotope activities:

(1) $^{238}$U series: $^{226}$Ra=1.0x10$^{-2}$ ± 1x10$^{-3}$ Bq kg$^{-1}$; $^{210}$Pb=1.74 ± 0.2 Bq kg$^{-1}$
$^{232}$Th series: $^{228}$Ra=6x10$^{-3}$ ± 2x10$^{-3}$ Bq kg$^{-1}$

(2) $^{238}$U series: $^{234}$U=4.24 x10$^{-2}$ ± 8x10$^{-4}$ Bq l$^{-1}$; $^{230}$Th=7.0x10$^{-5}$ ± 9x10$^{-6}$ Bq l$^{-1}$

(3) $^{238}$U series : $^{234}$U=1.2 1x10$^{-1}$ ± 1.7x10$^{-2}$ Bq kg$^{-1}$; $^{230}$Th=8.6x10$^{-3}$ ± 1.5x10$^{-3}$ Bq kg$^{-1}$

The activity concentrations of $^{238}$U and $^{232}$Th in concrete were measured through the detection of the third generation, short-lived descendants $^{234m}$Pa and $^{238}$Ac, respectively. The measurements for steel, water and wood confirm the absence of secular equilibrium in these materials. For steel it was deduced that the activity of $^{232}$Th is equal to that of its chemically identical decay isotope $^{228}$Th, while $^{238}$U activity is equal to that of $^{234}$Th – a direct descendent (2$^{nd}$ generation) with a half-life of 24 d, which is much smaller than the steel age (LSBB was built in the late seventies). In the cases of water and wood secular equilibrium was assumed at the measured activities of $^{232}$Th and $^{238}$U.

### 5. Neutron source

**5.1 Spontaneous fission**

The energy spectrum of neutrons emitted by the spontaneous fission of $^{238}$U and $^{232}$Th is described by a Watt formula: $e^{-E/a} \sinh(b\,E)^{1/2}$, with E the neutron energy, and a and b being radionuclide-specific fission parameters obtained from the SOURCES-4A data file [45] and E the neutron energy. Neutron multiplicities were similarly taken from Ref. 45, and the uncertainties were extracted from the original data compilation [46]. Spontaneous fission probability, half-life and atomic weights (the latter two required for the conversion between radionuclide concentration and activity) are obtained from the JEFF-3.2 library and the latest atomic mass evaluation [36, 47]. The data are listed in Table 3, including the neutron yield calculated on their basis.

**Table 3.** Spontaneous fission data and derived neutron yield (neutron per microgram of emitter and year). Uncertainties in atomic weight are $<5\times10^{-6}$ %.

|  | $^{238}$U | $^{232}$Th |
|---|---|---|
| Watt fission parameter *a* (MeV) | 0.6483 | 0.5934 |
| Watt fission parameter *b* (MeV$^{-1}$) | 6.811 | 8.030 |
| Atomic weight | 238.051 | 232.038 |
| Branching ratio, $p_{s.f.}$ | (5.46±0.1) x 10$^{-7}$ | (1.40±0.5) x 10$^{-11}$ |
| Neutron multiplicity, $\nu$ | 2.01±0.03 | 2.14±0.2 |
| Half-life, $T_{1/2}$ (y) | (4.46799±0.001) x 10$^{9}$ | (1.40503±0.006) x 10$^{10}$ |
| Neutron yield (n µg$^{-1}$ y$^{-1}$) | 0.430±2% | 3.84x10$^{-6}$±37% |

**5.2 Decay-induced neutrons**

Figure 3 shows the spectra of decay-induced neutrons in different materials of SIMPLE and Table 4 presents the neutron yield per microgram of neutron emitter in the material – both given by the USD code which assumes secular equilibrium in the decay chains. The yield of ($\alpha$,n) neutrons in BSG is very high, resulting from a significant production in boron: EXFOR data [36] shows that ($\alpha$,n) cross sections for $^{10}$B and $^{11}$B isotopes are on the order of 10$^2$ mb in the energy range of alphas emitted by the $^{238}$U and $^{232}$Th decay chains (~4-10 MeV). For comparison, Fig. 3 includes the BSG ($\alpha$,n) spectra retrieved by the improved version of SOURCES-4A employed by EDELWEISS that considers alpha energies up to 10 MeV [27].

For the GRP, we assume that the $^{238}$U and $^{232}$Th are essentially present in the glass fiber, which corresponds to neglecting their presence in the polypropylene matrix since organic materials generally have high radio-purity. In Table 4 the neutron yields corresponding to interactions in the surrounding gel matrix and inside the fiber were discriminated.

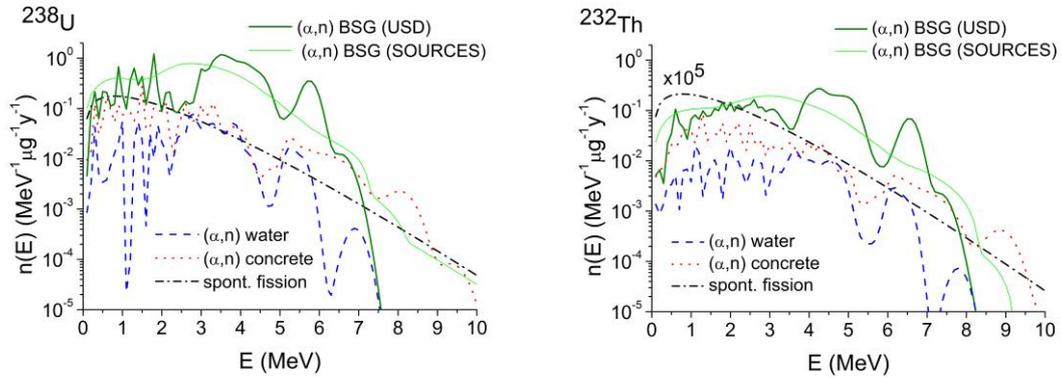

**Fig.3.** Spectra of ($\alpha$,n) neutrons due to the presence of $^{238}$U and $^{232}$Th (in secular equilibrium) for various materials. The Watt distribution representing spontaneous fission neutrons is also represented.

**Table 4.** Neutron yield (in neutron per microgram of emitter and year) from decay-induced ($\alpha$,n) reactions, due to the presence of $^{238}$U and $^{232}$Th in secular equilibrium for various materials.

|  | $^{238}$U | $^{232}$Th |
|---|---|---|
| Rock | 0.137 | 0.0458 |
| Concrete | 0.344 | 0.128 |
| Steel | 0.165 | 0.173 |
| Wood | 0.253 | 0.0766 |
| Water | 0.129 | 0.0403 |
| Paraffin | 0.322 | 0.0964 |
| Polyethylene | 0.324 | 0.0970 |
| Gel | 0.209 | 0.0640 |
| BSG | 2.34 | 0.660 |
| GRP [1] | 1.63 [0.331-1.74] | 0.506 [0.0994-0.557] |

[1] The lower and upper limits correspond to alpha interactions in the polypropylene and glass fiber, respectively. The adopted value considers a relative contribution of the polypropylene (glass fiber) to the GRP-induced event rate of 7.8% (92.2%) and 11.2% (88.8%) for ($\alpha$,n) reactions induced by the $^{238}$U and $^{232}$Th decay chains, respectively.

Interactions with the matrix occur with alphas emitted from the external layers of the roving, at a distance from the surface smaller than range of alphas in E-glass (14-40 μm for 4-8 MeV alphas, as calculated by the SRIM code [48]). Calculations with MCNPX [49] show that the fraction of alpha particles leaving the roving is (7.8±0.6)% and (11.2±0.8)% for $^{238}$U and $^{232}$Th in secular equilibrium, respectively. Only particles with energies > 450 keV were considered: this energy corresponds to the onset of EXFOR (α,n) cross sections in $^{13}$C which is the only isotope in the matrix yielding appreciable neutron production for the alpha energies involved. The uncertainties reported originate in the values obtained in the measurement of the roving diameter. On the basis of these results, the GRP signal includes the spontaneous fission component and a weighted average of (α,n) yields in the polypropylene matrix and glass fiber, the weights being 7.8% ($^{238}$U) and 11.2% ($^{232}$Th) for polypropylene and 92.2% ($^{238}$U) and 88.8% ($^{232}$Th) for E-glass.

While uncertainties of 18% are quoted for the SOURCES (α,n) yields based on a comparison with measurements [32], values corresponding to the USD data were not found in the literature. Additionally, the neutron spectra in Fig. 3 exhibit sharp peaks below ~3 MeV that are not displayed neither in measurements of e.g. Am-Be sources or angle-integrated alpha beams (where the convolution of the excited states forms only broad peaks [50, 51]) nor in the SOURCES-calculated spectra . Finally, the secular equilibrium assumption that usually applies to naturally occurring ores may become significantly altered when specific elements are extracted, either by human or natural processing. Typical examples already found (Table 2) are metals subject to smelting, as well as water and biological materials owing to the reduced solubility of Th in water. The information regarding the descendants is often incomplete, and a reasonable deduction of their activities becomes unfeasible. The accuracy of the USD yields and spectra, as well as the relevance of non-equilibrium to the uncertainty in the calculated event rates are evaluated in Section 8.1.

## 6. Neutron spectrum

Figure 4 shows the calculated neutron spectra for different configurations of GESA: spectrum in air prior to the installation of SIMPLE and on-detector spectra in Phase I and final stage of Phase II. Each spectrum generally consists of a fast neutron component from the source that is significantly degraded as neutrons scatter on their way to the tallying volume. In this regard, the spectra are essentially similar to that of a moderated nuclear fission reactor, and can be generically described as the combination of a fast Watt, a slowing-down epithermal 1/E and a thermal Maxwell distribution. Table 5 summarizes the results concerning thermal (< 0.5 eV) and fast (> 1 MeV) neutron fluence rates in the evaluated configurations.

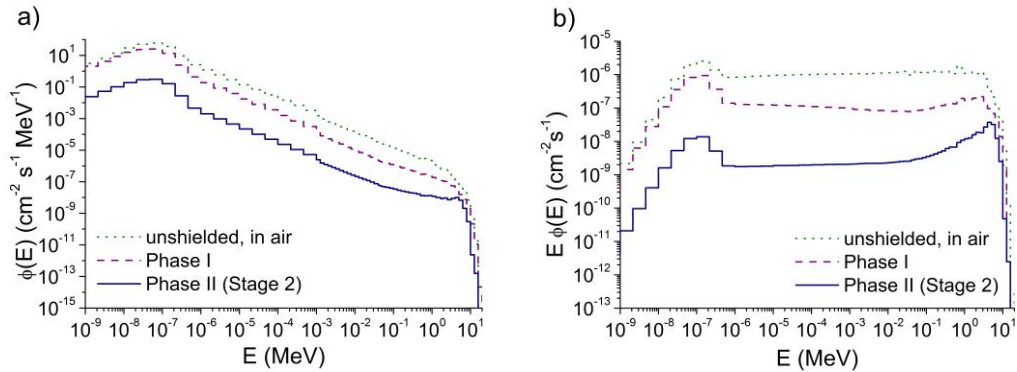

**Fig. 4.** Calculated neutron spectra in GESA for the various development phases of SIMPLE, in (a) the conventional representation per unit energy and (b) the representation per unit lethargy where the thermal and fast neutron components are displayed more clearly.

**Table 5.** Thermal and fast neutron fluence rates for various configurations of GESA.

|  | $\phi_{thermal}$ (cm$^{-2}$ s$^{-1}$) | $\phi_{fast}$ (cm$^{-2}$ s$^{-1}$) |
|---|---|---|
| unshielded, in air | 5.92x10$^{-6}$ | 1.63x10$^{-6}$ |
| Phase I | 2.12x10$^{-6}$ | 3.03x10$^{-7}$ |
| Phase II (Stage 2) | 3.08x10$^{-8}$ | 4.30x10$^{-8}$ |

The results show that the neutron field intensity decreases 2 orders of magnitude as neutrons traverse the increased moderator thickness used in Phase II. The modification in the neutron spectra caused by the shielding is most evident if one compares the maxima in the thermal and fast regions using the representation per lethargy (Fig. 4b) commonly used in reactor physics. Relative to the unshielded configuration (where $\phi_{fast}/\phi_{thermal}$=0.28), there is a noticeable reduction of the fast-to-thermal component in Phase I ($\phi_{fast}/\phi_{thermal}$=0.14) due to the small moderator consisting of the SDD gel matrix and tank water. The situation is reversed for the heavily filtered spectrum in Phase II ($\phi_{fast}/\phi_{thermal}$=1.4), which is hardened as only the highest-energy neutrons can penetrate through the shielding.

Figure 5 provides further insight into the neutron spectrum for Phase II, with a discrimination of the contributions of various materials and reactions to the overall neutron fluence rate. The water shield effectively reduces the neutron background sources to detector components, the detector container being clearly the major contributor due to the high yield of ($\alpha$,n) reactions in BSG. In order to improve the accuracy of event rate calculations, a correction of -0.7% is applied to the intensity of the BSG source term that accounts for the mismatch between the calculated and measured container masses.

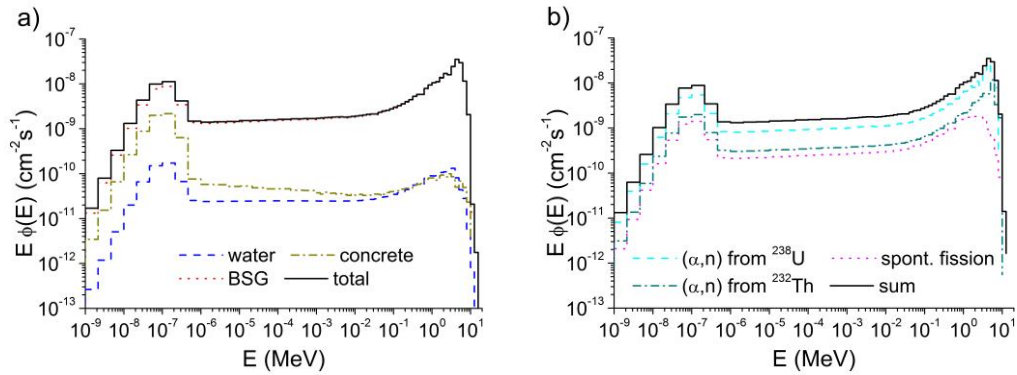

**Fig. 5.** Calculated on-detector neutron spectra for Stage 2 of SIMPLE Phase II with a discrimination of a) selected material contributions and b) neutron-producing reactions in BSG.

## 7. Event rates

### 7.1 Reactions

The $E_{crit}$ and $LET_{crit}$ thresholds referred to in Section 1 determine the minimum nuclear recoil energy ($E_{rec}$) for the production of a bubble nucleation. For the particles of interest (neutrons and nuclear recoils) the LET can be considered equal to the stopping power, i.e. *bremsstrahlung* radiation and secondary electrons (delta-rays) are neglected. Stopping powers calculated by SRIM were used to extract $E_{rec}$.

Neutron scattering (elastic and inelastic) were scrutinized as event-producing sources, as well as transmutation reactions with positive Q-values: (n,p) and (n,$\alpha$) in $^{35}$Cl. The relation between nuclear recoil and neutron energies is defined by standard kinematic equations [52] involving the reaction Q-value, particle atomic weights and the angular deviation upon interaction. For each reaction, the minimum neutron energy yielding a nucleation ($E_{min}$) was derived for head-on collisions corresponding to the maximum energy transfer.

The values of $E_{rec}$ and $E_{min}$ are presented in Table 6, for the various reactions considered. For inelastic scattering (not included in previous background estimates), the data refers to the first excited state, which has smaller |Q| hence is more populated.

Heavy recoiling nuclei (F, Cl, S, P) have high LET, and the constraint in $E_{rec}$ is determined by $E_{crit}$; in contrast, for lighter particles such as C nuclei, $E_{rec}$ is set by $LET_{crit}$.

The transmutation reactions in $^{35}$Cl can induce a nucleation regardless of neutron energy, because Q > 0 and the heavier reaction products emerge with energy larger than 17 keV ($^{35}$S) and 104 keV ($^{32}$P). A similar effect occurs for the recoiling nuclei from inelastic scattering in C and Cl (for neutron energies above the reaction threshold, since Q < 0) which emerge with minimum energies of 344 keV ($^{12}$C), 34 keV ($^{35}$Cl) and 46 keV ($^{37}$Cl); the minimum neutron energies in Table 6 are therefore equal to the reaction thresholds. The recoiling nucleus from inelastic scattering in F emerges with a minimum energy of 5.5 keV which is lower than the $E_{rec}$; in this case $E_{min}$ is larger than the reaction threshold energy.

**Table 6.** Minimum nuclear recoil ($E_{rec}$) and neutron ($E_{min}$) energies for the production of a nucleation in $C_2ClF_5$.

| Reaction | Q-value (keV) | $E_{rec}$ (keV) | $E_{min}$ (keV) | Isotopic Abundance (%) |
|---|---|---|---|---|
| *Elastic scattering* | | | | |
| $^{12}$C(n,n)$^{12}$C | 0 | 115 | 402 | 98.93±0.088 |
| $^{13}$C(n,n)$^{13}$C | | 123 | 460 | 1.07±0.08 |
| $^{19}$F(n,n)$^{19}$F | | 8 | 43 | 100 |
| $^{35}$Cl(n,n)$^{35}$Cl | | 8 | 75 | 75.76±0.10 |
| $^{37}$Cl(n,n)$^{37}$Cl | | 8 | 79 | 24.24±0.10 |
| *Inelastic scattering* | | | | |
| $^{12}$C(n, n$_1$')$^{12}$C | -4438.91±0.31 | 115 | 4812 | 98.93±0.088 |
| $^{13}$C(n, n$_1$')$^{13}$C | -3089.443±0.020 | 123 | 3329 | 1.07±0.08 |
| $^{19}$F(n, n$_1$')$^{19}$F | -109.894±0.05 | 8 | 120 | 100 |
| $^{35}$Cl(n, n$_1$')$^{35}$Cl | -1219.29±0.11 | 8 | 1255 | 75.76±0.10 |
| $^{37}$Cl(n, n$_1$')$^{37}$Cl | -1726.58±0.04 | 8 | 1774 | 24.24±0.10 |
| *Transmutation reactions with Q>0* | | | | |
| $^{35}$Cl(n,p)$^{35}$S | 615.0234±0.0532 | $^{35}$S:8 | 0 | 75.76±0.10 |
| $^{35}$Cl(n,$\alpha$)$^{32}$P | 937.74±0.05 | $^{32}$P: 8; $\alpha$: 269-1981 | 0 | 75.76±0.10 |

**7.2 Cross sections**

Table 6 displays differences between the isotopes of C and Cl with respect to $E_{min}$. For the event rate calculations some simplifications were introduced based on the reaction cross sections, with a selection shown in Fig. 6.

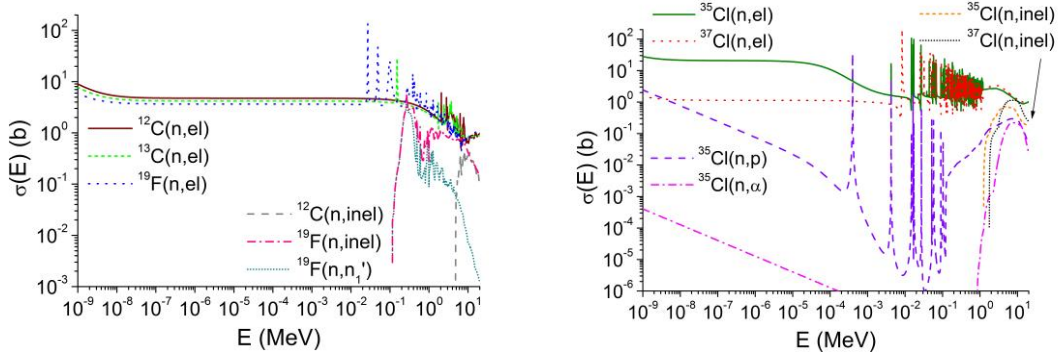

**Fig. 6.** Cross sections of selected neutron reactions originating events in SIMPLE SDDs. The data are taken from the ENDF-B7.1 library, except for $^{12}$C (CENDL-3.1) and $^{13}$C (JEFF-3.2) [34]. Labels (n,el) and (n,inel) refer to total elastic and total inelastic scattering, respectively.

The contribution of elastic scattering on $^{13}$C in the energy interval 402-460 keV has not been discriminated because this is a minor isotope with slightly smaller interaction cross section than that of $^{12}$C; the cross section of natural C was used above 402 keV. The cross section of natural Cl was used beyond 75 keV to describe elastic scattering, as its two isotopes have similar energy thresholds and the contribution of $^{37}$Cl is simultaneously decreased since it is the less abundant and has the smaller cross section.

The data in Table 6 for inelastic scattering refer to the first excited state. Although total inelastic cross sections are significantly larger, the states of higher orders have increased $E_{min}$ yielding the corresponding event rates smaller than inferred by cross section ratios. The total inelastic cross section starts to deviate from the one for the first excited state at about 2 MeV (Cl), 5 MeV (C) and 10 MeV (F) where SDDs have a decreased neutron population. As a first approximation, inelastic scattering was dealt with by considering $E_{min}$ for the first excited state and the total inelastic cross section, thereby providing an upper limit to its contribution.

### 7.3 Conversion of neutron fluence to event rates

The rate (R) per target atom of each nucleation-inducing reaction is generically calculated as $R = \int \sigma(E)\phi(E)dE$. With MCNP output group fluence rates, the event rate (R) due to each reaction can be calculated as a sum over the various energy bins:

$$R = \sum_i N \langle\sigma\rangle_i \phi_i$$

where N is the number of target atoms, $\phi$ is the calculated group neutron fluence rate and $\langle\sigma\rangle_i$ is the spectral-averaged reaction cross section in the i$^{th}$ energy bin.

The average group cross sections were calculated using FLXPRO. The original pointwise-evaluated cross section data was initially converted to the standard SAND-II 640 group structure (45 equi-lethargy groups per decade from 1 meV to 1 MeV and group width of 100 keV for 1 - 20 MeV) [53] with a flat weighting spectrum. These fine group cross sections were later converted to the group structure of the simulations using a reactor weighting spectrum consisting of a Maxwell, 1/E and Watt distributions.

Differential and group energy distribution of the recoils were calculated for elastic and inelastic scattering, the only event-producing reactions available within SPECTER. The recoil spectrum for each reaction was normalized to the respective R, and folded with the nucleation efficiency in order to determine the number of recoils with energies larger than $E_{min}$. The nucleation efficiency was described as $\eta(E) = 1 - \exp[-\Gamma(E/E_{rec} - 1)]$ for $E \geq E_{rec}$ (and 0 otherwise) with $\Gamma=4.2\pm0.3$ [1]. Group efficiency values were calculated with FLXPRO using a cubic splines interpolation of the differential recoil distribution in the SAND-II bins as weighting spectrum.

The reaction rates of the two transmutation reactions were added to the event rate because all recoils fulfill the threshold conditions (translated as $E_{min}=0$ in Table 6).

## 8. Results

Table 7 shows the calculated event rates, with a discrimination of contributions of the various materials and neutron-producing reactions. In the case of GRP, values corresponding to (α,n) interactions in the polypropylene matrix (lower value) and in the glass fiber are given; a weighted average is reported in the sum considering that the relative contribution of the alpha interactions occur within the glass fiber (Section 5.2). Various results correspond to upper limits that reflect the experimental detection limits regarding the radio-assays. With the concrete shielding neutrons from the rock, the upper limits reported for the latter reflect the use of the shallowest concrete layer in the simulations. Due to its negligible contributions to the total event rate, these upper limits are not subject to more refined analysis.

The results are organized in 3 groups: (i) the structural materials of the facility; (ii) materials added to shield against neutrons; (iii) materials used to set-up the detector. In contrast to the structural materials, both the external shield and the detector materials may be subject to modification in future SIMPLE developments.

**Table 7.** Calculated event rates, discriminating the neutron source materials and reactions.

| Neutron origin | Event rate (evt/kg-d) | | | |
|---|---|---|---|---|
| | s.f. | $(\alpha,n)$ from $^{238}$U | $(\alpha,n)$ from $^{232}$Th | Sum |
| *Structural materials* | | | | |
| Rock | <$4.26 \times 10^{-7}$ | <$2.69 \times 10^{-6}$ | <$6.33 \times 10^{-9}$ | <$3.12 \times 10^{-6}$ |
| Concrete | $8.46 \times 10^{-4}$ | $1.23 \times 10^{-3}$ | $9.67 \times 10^{-4}$ | $3.04 \times 10^{-3}$ |
| Steel | $1.08 \times 10^{-8}$ | $2.30 \times 10^{-9}$ | $3.55 \times 10^{-9}$ | $1.67 \times 10^{-8}$ |
| Wood | $1.69 \times 10^{-6}$ | $4.74 \times 10^{-6}$ | $1.11 \times 10^{-7}$ | $6.54 \times 10^{-6}$ |
| *Shield* | | | | |
| Water-external shield | $2.92 \times 10^{-4}$ | $1.58 \times 10^{-4}$ | $2.47 \times 10^{-7}$ | $4.49 \times 10^{-4}$ |
| Paraffin | <$3.62 \times 10^{-8}$ | <$3.23 \times 10^{-7}$ | <$4.43 \times 10^{-7}$ | <$8.02 \times 10^{-7}$ |
| Polyethylene | $1.36 \times 10^{-5}$ | $5.30 \times 10^{-5}$ | $4.25 \times 10^{-5}$ | $1.09 \times 10^{-4}$ |
| *Intrinsic* | | | | |
| GRP [1] | $2.65 \times 10^{-3}$ | [$4.14 \times 10^{-3}$ - $1.22 \times 10^{-2}$] | [$6.77 \times 10^{-3}$ - $3.07 \times 10^{-2}$] | $4.37 \times 10^{-2}$ |
| Water-tank | $1.28 \times 10^{-3}$ | $5.55 \times 10^{-4}$ | $8.36 \times 10^{-7}$ | $1.83 \times 10^{-3}$ |
| BSG | $3.46 \times 10^{-2}$ | $2.07 \times 10^{-1}$ | $8.07 \times 10^{-2}$ | $3.23 \times 10^{-1}$ |
| Gel | <$3.54 \times 10^{-4}$ | <$1.44 \times 10^{-4}$ | - | <$4.97 \times 10^{-4}$ |
| Microphone | <$1.74 \times 10^{-4}$ | <$6.68 \times 10^{-5}$ | - | <$2.41 \times 10^{-4}$ |
| | | | Total | $3.72 \times 10^{-1}$ |

[1] The lower and upper limits correspond to alpha interactions in the polypropylene and in the glass fiber, respectively. The adopted value in the last column considers a relative contribution of the polypropylene (glass fiber) to the GRP-induced event rate of 7.8% (92.2%) and 11.2% (88.8%) for $(\alpha,n)$ reactions induced by the $^{238}$U and $^{232}$Th decay chains, respectively.

The results suggest that most of the detector background signal is created by the detector container of BSG (87%) and to a lesser degree by the tank GRP walls (12%). The contribution of the structural materials (mostly from the concrete) is reduced to 0.8% and the external shielding (essentially the water) contributes with less than 0.2% to the background.

Figure 7 shows the energy distribution of the recoils; the integral above the relevant thresholds yields the calculated event rates. These are presented in Table 8, with a discrimination of the event-inducing reactions in $C_2ClF_5$.

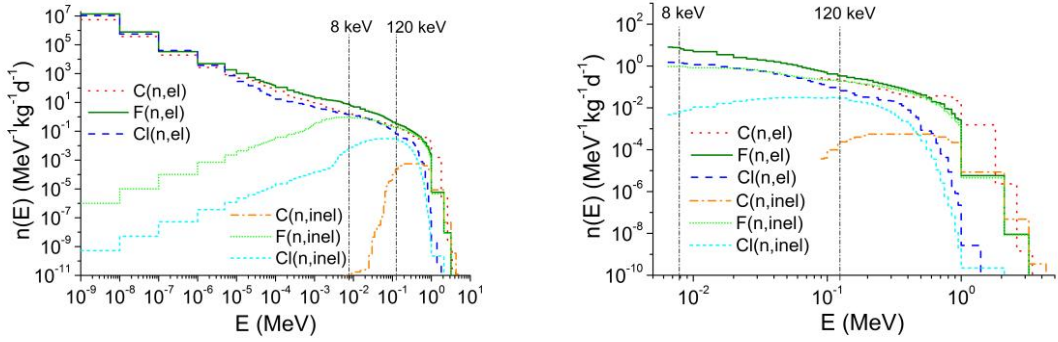

**Fig. 7.** Energy distribution of the recoil nuclei with a zoom over the region above threshold. The vertical lines locate the $E_{rec}$ thresholds for Cl and F (8 keV), as well as C (120 keV), below which no events are induced (in the zoom view the recoil distributions below threshold are deleted).

**Table 8.** Contribution of the various event-producing reactions to the detector signal. The statistical uncertainty for each reaction varies between 0.6% and 1.1%.

| Reaction | Event rate (evt/kg-d) |
|---|---|
| C(n,el) | $3.71 \times 10^{-2}$ |
| Cl(n,el) | $4.25 \times 10^{-2}$ |
| F(n,el) | $2.10 \times 10^{-1}$ |
| $^{35}$Cl(n,p) $^{35}$S | $5.42 \times 10^{-3}$ |
| $^{35}$Cl(n,$\alpha$) $^{32}$P | $4.33 \times 10^{-4}$ |
| C(n,inel) | $3.83 \times 10^{-4}$ |
| F(n,inel) | $6.98 \times 10^{-2}$ |
| Cl(n,inel) | $6.44 \times 10^{-3}$ |
| Total | $3.72 \times 10^{-1}$ |

There is a clear predominance (56%) of elastic scattering in fluorine that results from its large abundance in the superheated liquid, and a residual contribution (1%) of the transmutation reactions in the Cl isotopes due to their reduced cross sections.

With inelastic scattering in F corresponding to 19% of the detector signal, evaluation of the resulting error by the simplified treatment (Section 6.2) is pertinent. An accurate calculation was performed for fluorine considering 21 excited states and the total neutron spectrum. The contribution of the excited states to the total F(n,inel) event rate is 29% ($n_1'$), 49% ($n_2'$), 6% ($n_3'$ and $n_4'$), 9% ($n_5'$), 1% ($n_6'$) and less than 0.3% for each of the higher order states. This detailed calculation yielded 98.3% of the value previously determined, to which a correction of -1.7% is therefore applied.

Relative to the previously reported event rate of 0.33 evt/kgd [1, 32], the inclusion of inelastic scattering reactions and the consideration of the recoil energy distribution introduced corrections of +26% and -23% respectively. These nearly compensate each other, yielding a net variation of +13%, essentially due to the tank wall contribution.

## 8.1 Uncertainty analysis

The statistical uncertainties in the group fluence rates were kept below 10% for each bin by simulating a sufficiently large number of source particles and employing variance reduction techniques available in MCNP when necessary. The propagated statistical uncertainty in the event rate is 0.51%. Non-statistical uncertainties related to material radio-assays have been indicated previously. We now address the missing contributions to the total uncertainty.

### a) ($\alpha$,n) neutron production yield

The uncertainty in ($\alpha$,n) yields provided by the USD code is estimated by a comparison with experimental thick target yields [51, 54, 55]. The measurements from Ref. 55 cover the $^{238}$U and most of the $^{232}$Th alpha emission spectra and have been generally adopted for the elemental yields. Otherwise, theoretical results [56, 57] normalized to the experimental data are used to extend its energy range. Cubic splines interpolation and linear extrapolation are applied – the latter for the 10.18 MeV and 11.66 MeV alphas from $^{212}$Po thereby creating an uncertainty in the ($\alpha$,n) "experimental" yield for the $^{232}$Th spectrum. The yields for the mixtures and compound materials that contribute most to the detector signal (BSG/E-glass, water and concrete) were calculated within ±9% [58] using stopping power ratios and the method outlined in Ref. 59; theoretical elemental yields were used for the constituents for which experimental values were not available.

For the $^{238}$U decay chain spectrum, the following ratios of measured to calculated (M/C) yields were obtained: 1.19 (B), 1.13 (C), 1.46 (O), 0.95 (Al), 1.17 (Si), 1.10 (Fe) and 1.22 (BSG), 1.23 (water), 1.20 (concrete).

In the case of $^{232}$Th decay chain spectrum, the extrapolated contribution of the high energy $^{210}$Po alphas to the experimental elemental yields ranges from 27% (for B and C) to 75% (for Fe). For compounds or mixtures, M/C values of 0.66±0.14 (BSG), 0.68±0.25 (water) and 0.74±0.14 (concrete) were obtained. The quoted uncertainties are the propagation of those in the elemental neutron yields, being merely indicative. When parabolic end conditions are used for the extrapolation (similarly to Ref. 58), the results are consistent with M/C=1±0.15.

The USD yields were further compared with the values calculated by SOURCES-4A (Section 5.2). The SOURCES/USD ratios were 1.35 (BSG), 1.21 (water), 1.36 (concrete) in the $^{238}$U decay spectrum and 1.72 (BSG), 1.52 (water), 1.71 (concrete) in the $^{232}$Th decay spectrum. In order to obtain consistent values among the various data sets (within the 18% uncertainty quoted for SOURCES), uncertainties of 20% and 45% are assigned to the USD yields for the $^{238}$U and $^{232}$Th decay spectra, respectively.

**b) ($\alpha$,n) neutron spectrum**

Possible inaccuracies in the input spectra (namely their sharp peaks) were assessed by calculating the BSG contribution to the detector signal with the source ($\alpha$,n) spectrum from SOURCES-4A. Results were normalized to the USD ($\alpha$,n) yield for the measured concentrations of $^{238}$U and $^{232}$Th in BSG. The calculated on-detector neutron spectra and the 90% response regions (i.e., the 5% and 95% limits) as well as the median for the various event-producing reactions are shown in Fig. 8. The output bins have the same width as the energy difference of the source spectra shown in Fig. 3. The statistical uncertainty in the fluence rate is <0.1% in each bin.

Figure 8 shows that the narrow features in the USD input spectrum spread out as neutrons are scattered by the hydrogenous gel. Above ~2 MeV the on-detector neutron energy distribution using the USD data displays the superposition of the broad peaks in the $^{238}$U and $^{232}$Th ($\alpha$,n) spectra (Fig. 3), being therefore shifted to higher energies than that using SOURCES-4A.

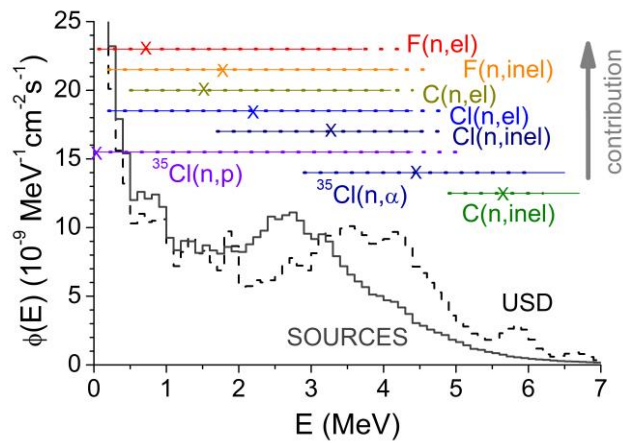

**Fig. 8.** Calculated on-detector neutron spectra induced by ($\alpha$,n) reactions in BSG for source spectra obtained from the USD and SOURCES-4A codes. The 90% response regions (solid line: SOURCES; dotted: USD) for the event-producing reactions are ordered according to their relative contribution to the detector signal (Table 7); the median (X) refers to the USD spectrum.

When the contributions of the various ($\alpha$,n) reactions is discriminated, reactions with more than 50% response below ~3 MeV had USD/SOURCES response ratios of 1.03-1.04 except for elastic scattering in F (0.95) and C (1.09). Beyond this energy discrepancies increase from 30% (inelastic scattering in Cl) up to 150% (inelastic scattering in C) reflecting the different on-detector integral neutron fluence rates in the response regions using the USD and SOURCES distributions. However, these are minor contributors (<1.5%) to the detector signal that do not affect significantly the results of the comparison. The overall deviation in the USD calculated contributions to the event rate from BSG relative to those using SOURCES input spectra is -0.52% from ($\alpha$,n) reactions, being further reduced to -0.46% when spontaneous fission is included.

### c) Secular equilibrium

Non-equilibrium was evident for some materials involved in the experiment and is likely to occur also in BSG (and in the E-glass of GRP). A point of concern is the $^{226}$Ra sub-chain (half-life of 1600 y) in the decay of $^{238}$U. However, the evaluation of various glasses - including BSG from the producer of SIMPLE detector containers - shows no evidence of disequilibrium at this point of the chain [60, 61]. Still, as a result of the mixing process at 1600 $^{\circ}$C [62] oxides of the $^{238}$U and $^{232}$Th descendants may be released. The only possible cause of relevant disequilibrium is PbO with a boiling point of 1470 $^{\circ}$C [63]. Lack of secular equilibrium in the $^{238}$U chain due to a depletion of $^{210}$Pb (half-life 22 y) yields an uncertainty of 12%, corresponding to the relative contribution of the $^{210}$Pb sub-chain.

### d) Cross section data

The uncertainty associated to the reaction cross sections was evaluated using the FLXPRO code to convert their uncertainties (variances and co-variances) from the originally evaluated data to the group structure of the calculated neutron spectrum. The analysis was applied to the principal contributing reactions, yielding relative uncertainties of 1.8% and 3.1% for elastic and inelastic scattering in F and 0.34% for elastic scattering on C.

The potential error introduced by the application of group cross sections was evaluated by a comparative calculation of the detector signal via convolution of the continuous-energy reaction cross sections prepared for MCNP with the neutron fluence rate during the simulation runs (FM tally multiplier card of MCNP). Discrepancies smaller than 0.3%±0.6% (stat.) were found for all event-producing reactions except for those reactions responding mostly in the high energy region: $^{35}$Cl(n,$\alpha$) and inelastic scattering (Fig. 6), with discrepancies ranging from 0.7% to 2.5% (±1.1% stat.) that could benefit from a finer group structure. An overall discrepancy of 0.31% ± 0.51% (stat.) was found that is therefore neglected in the uncertainty analysis.

### e) Inelastic scattering

Inelastic scatterings in C and Cl have modest contributions to the detector signal (0.1% for C; 2% for Cl): their simplified treatment is not subject to finer analyses. Based on the correction of 1.7% calculated for F, we attribute a conservative uncertainty of 5% to the calculated values for C and Cl.

### f) Material composition

BSG and GRP have important contributions to the background signal due to their high boron content in combination with considerable amounts of $\alpha$-emitting radio-impurities. The measurement uncertainty for the B content in BSG is 4%, in the case of GRP, for which a nominal composition was used, an uncertainty of 10% (1-$\sigma$) is considered reflecting the range of variability in the reference data. The uncertainty of the weight factors, attributed to the contributions of (n,$\alpha$) reactions in the E-glass and the embedding polypropylene, is estimated as 7.2% and represents the uncertainty in the roving diameter.

### g) Nucleation efficiency

The uncertainty associated with the $\Gamma$ factor employed in the efficiency curve was determined by repeating the calculations for elastic and inelastic scattering in F with the extreme values of 3.9 and 4.5 (Section 7.3). The variation in the sum of the corresponding event rates was 0.6%.

### h) Simulation

Any simulation code has inherent inaccuracies due to, for example the transport cross sections and the interaction models. In addition, there are unavoidable errors in the geometry description or material composition of such a complex and large set-up such as SIMPLE. This is particularly relevant for neutron transport, being strongly influenced by the interacting elements and isotopes. As shown in previous studies, e.g. Refs. 64, 65, discrepancies of 10% between absolute neutron simulation and experiment are common. This value is adopted as an estimate of the uncertainty associated with the simulation procedure.

### i) Muon-induced neutrons

The interaction of cosmic muons with rock generates high energy neutrons that emerge from the cavern surface and contribute to the detector signal. At GESA's depth, the estimated muon-induced neutron fluence rate is $4.0 \times 10^{-8}$ n cm$^{-2}$ s$^{-1}$ and the average muon energy is 177 GeV [66]. Both the total muon interaction cross section as a function of energy and the media dependence of neutron production are generally described by a power law, in the latter case involving the product of the rock density ($\rho$) and its average atomic number <A> ($\rho$=2.61 g cm$^{-3}$ and <A>=23.4 for GESA).

The contribution of muon-induced neutrons to the detector signal was estimated using MCNPX, considering a vertical muon incidence that induces a surface neutron source at GESA's ceiling with the angular distribution given by Ref. 66. The same reference reports group neutron fluence rates (<1MeV, 1-10 MeV, 10-100 MeV and > 100 MeV) for various sites, that were used to normalize the simulations of GESA. The relative contribution of each group to the total fluence rate was fit (power law) to the $\rho$ <A> and afterwards to the average muon energy of the site, in order to derive within 20% the 4-group source fluence rates. The neutron energy distribution in the 2 groups above 10 MeV was extracted from Ref. 67, with distribution parameters calculated for the average muon energy. For the 2 groups below 10 MeV, flat distributions were used.

Reaction cross sections from JENDL-HE-2007 [36] were used to calculate the event rates. As shown in Table 9, the estimated muon-induced event rate is $2.1 \times 10^{-3}$ evt/kgd, corresponding to 0.56% of the total value.

**Table 9.** Estimated muon-induced recoil event rates. Monoenergetic neutrons with the maximum group energy were considered at ≤10MeV (two lowest energy groups).

| Neutron energy | Fluence rate ($10^{-9}$ cm$^{-2}$s$^{-1}$) | Event rate ($10^{-4}$ evt/kgd) |
|---|---|---|
| 0 – 1 MeV | 28 | ~0 |
| 1 – 10 MeV | 1.8 | 0.11 |
| 10 – 100 MeV | 7.6 | 8.4 |
| 0.1 – 3 GeV | 2.6 | 12 |

### j) Photon-induced nucleations

The liquid sensitivity to radiation is characterized by the reduced superheat factor $S=(T-T_b)/(T-0.9T_c)$ where $T$, $T_b$ and $T_c$ are the operating, boiling and critical temperatures at the operating pressure [68]. Numerous studies have shown the insensitivity of various liquid devices to minimum ionizing radiation (photons, muons, electrons, etc.) when S<0.52, with rejection factors well below $10^{-9}$ [3, 4, 68, 69]. SIMPLE SDDs run at S=0.34; although an insensitivity to photons has been observed [70], neither the rejection factor nor the environmental background in the dark matter search experiment have been measured. When the limits from COUPP [3] are extrapolated to $C_2ClF_5$ and to a rejection efficiency <$10^{-9}$, we estimate less than $10^{-3}$ evts/kgd in SIMPLE from low energy photons and beta decays. Nucleations induced by photonuclear reactions were also considered for PICO; a preliminary upper limit of ~$10^{-2}$ evt/kgd was derived upon the no-observation of photons with energies larger than 10 MeV at the time of report that is subject to improvement with extended measurement time [4, 21].

k) **Multiple scattering**

The time resolution of the signal acquisition system of 80 $\mu$s restricts the capability to reject multiple hits in the same SDD or in different SDDs. From the MCNP simulations, the number of events per neutron entering the detector volume is 0.6%. Considering the neutron energy loss after interaction (with the target and matrix atoms), multiple hits within a detector induce therefore less than 0.6% of the measured recoil rate - a value that is further decreased in the case of multiple interactions in different SDDs.

In Table 10 an uncertainty analysis is presented. Uncertainties of 100% have been assigned to small quantities lacking a reliable uncertainty estimate. The sensitivity coefficients (i.e., the relative contribution of each quantity to the total event rate) is included to derive the propagated uncertainty. Only values yielding relative uncertainties larger than 0.1% are shown.

The results show that the largest identified uncertainties are associated with the ($\alpha$,n) yields, the deviations from secular equilibrium, the measurement uncertainty regarding $^{238}$U in BSG and the simulation procedure – their combination alone yielding essentially the overall uncertainty of 22.0%.

The final estimate for the calculated background event rate in Stage 2 of SIMPLE is 0.372 $\pm$ 0.002(stat.) $\pm$ 0.097 (non-stat.) evt/kgd. Relatively to previous estimates, the non-statistical uncertainty is increased from 11.5% to 22.0%. The uncertainties associated with the ($\alpha$,n) yields and the simulation itself set a minimum value of 18% for the non-statistical uncertainty; the measurement uncertainty of U in the BSG and the non-equilibrium in BSG contribute each with a supplementary 1-2% yielding basically the total non-systematic uncertainty.

**Table 10.** Uncertainties (1-σ) in the calculated background event rate. Only values yielding propagated uncertainties larger than 0.1% are presented.

| Quantity | Uncertainty In quantity (%) | Origin of uncertainty | Sensitivity coefficient (%) | Propagated uncertainty (%) |
|---|---|---|---|---|
| Statistical | 0.51 | simulation statistics | 100 | 0.5 |
| (α,n) yield | $^{238}$U: 20 | comparison with | 59.4 | 11.9 |
| | $^{232}$Th: 45 | experimental data | 29.5 | 13.3 |
| (α,n) spectrum | 0.52 | sharp structures | 89.2 | 0.4 |
| (α,n) yield - U in BSG and GRP | 12 | non-equilibrium | 59.1 | 7.1 |
| (α,n) yield – U and Th in water | 100 | | 0.19 | 0.2 |
| F(n,el+inel) react. rate | 0.57 | nucleation efficiency | 75.2 | 0.4 |
| F(n,el) react. rate | 1.8 | cross section data | 56.5 | 1.0 |
| F(n,inel) react. rate | 3.1 | | 18.8 | 0.6 |
| | 1.7 | simplified calculation | 18.8 | 0.3 |
| U in BSG | 15.0 | measurement | 65.0 | 9.7 |
| Th in BSG | 8.7 | | 21.7 | 1.9 |
| U in GRP | 6.3 | | 4.1 | 0.3 |
| Th in GRP | 10.6 | | 7.5 | 0.8 |
| B in BSG | 4.0 | | 77.4 | 3.1 |
| B in GRP | 10.0 | nominal composition | 10.9 | 1.1 |
| MCNP output | 10 | model inaccuracies | 100 | 10 |
| Muon-induced neutrons | 100 | estimated contribution | 0.56 | 0.6 |
| Photon nucleations | 100 | | 0.3 | 0.3 |
| Multiple scattering | 100 | | 0.6 | 0.6 |
| | | | Combined relative uncertainty | 24.7 |

## 9. Conclusions

The estimated SDD neutron background signal for Stage 2 of SIMPLE Phase II was simulated using the MCNP code, yielding $0.372 \pm 0.002$(stat.) $\pm 0.097$ (non-stat.) evt/kgd. For the net exposure of 6.71 kgd in Stage 2 this corresponds to 2 estimated events vs. 1 measured.

This final estimate of the recoil background included various additional items relative to previous works, both in the event rate calculation and in the assessment of non-statistical uncertainties. Differences up to 26% in the event rate were seen for individual contributions. However, some counterbalanced each other yielding a net increase of 13%. This value corresponds to an error in the previously anticipated recoil background estimate caused by the introduction of various simplifications (e.g. the treatment of recoil energy distributions, consideration of inelastic scattering and additional materials of the experiment).

The non-statistical uncertainty component increased from 12% to 25% as a result of an evaluation of the accuracy in the ($\alpha$,n) yields employed, and inclusion of uncertainties due to the simulation process and non-equilibrium in the decay chains. The first two set a limit of 21% on the non-statistical uncertainty that cannot be decreased with improved characterization of the experiment materials.

The revised estimate of the detector background signal changes insignificantly the Feldman-Cousins analysis of its contribution to the overall Phase II measurements.

Next developments in SIMPLE [71] depend on reducing the neutron background using boron-free materials and increased radio-purity. Promising candidates that have been identified should be assayed using techniques with lower detection limits (ICPMS, low-level spectroscopy). With a reduction of the intrinsic background by two orders of magnitude relatively to Phase II, muon-induced neutrons are expected to become important contributors to the SDD signal. The application of FLUKA or GEANT4 is anticipated in order to assess this background component in future SIMPLE experiments and maintain uncertainties below 30-40% [72].

## 10. Acknowledgements

We thank Michel Auguste, Daniel Boyer, Alain Cavaillou and Christophe Sudre (LSBB) for their assistance in the Stage 2 shielding, and provision of the additional materials samples assayed in this report. We also thank Harry Miley and Rosey Payne (PNNL), Pia Loaiza (LSM) and Christophe Destouches (Univ. Avignon) for the radio- and chemical assays of the GESA construction materials.

We are grateful to Larry Greenwood (Argonne National Laboratory) for providing the full version of the SPECTER code, and to Vitaly Kudryavtsev (University of Sheffield) for offering the improved version of SOURCES-4A. Luís Alves ($C^2$TN) is acknowledged for the microscope usage.

This work was supported by Grants PDTC/FIS/115733/2009, PDTC/FIS/121130/2010, IF/00628/2012/CP0171/CT0008, SFRH/BPD/94028/2013 and UID/Multi/04349/2013 of the Portuguese Foundation for Science and Technology.

## 11. References

[1] M. Felizardo, T. A. Girard, T. Morlat et al., The SIMPLE Phase II dark matter search, *Physical Review D* **89** (2014) 072013. doi: 10.1103/PhysRevD.89.072013

[2] S. Archambault, E. Behnke, P. Bhattacharjee et al., Constraints on Low-Mass WIMP Interactions on $^{19}$F from PICASSO, *Physics Letters B* **711** (2012) 153-161. doi:10.1016/j.physletb.2012.03.078

[3] E. Behnke, J. Behnke, S. J. Brice et al., First dark matter search results from a 4-kg CF3I bubble chamber operated in a deep underground site, *Physical Review D* **86** (2012) 052001. doi: 10.1103/PhysRevD.86.052001

[4] C. Amole, M. Ardid, D.M. Asner et al., Dark matter search results from the PICO-2L C3F8 bubble chamber. *Physical Review Letters* **114** (2015) 231302. doi:10.1103/PhysRevLett.114.231302

[5] R.E. Apfel, The superheated drop detector, *Nuclear Instruments and Methods* **162** (1979) 603-608.

[6] F. Seitz, On the theory of the bubble chamber, *Physics of Fluids* **1** (1958) 2-13.

[7] M. Felizardo, R.C. Martins, A.R. Ramos et al., New acoustic instrumentation for the SIMPLE superheated droplet detector, *Nuclear Instruments and Methods* **A 589** (2008) 72-84. doi:10.1016/j.nima.2008.02.012

[8] M. Felizardo, T. Morlat, A. C. Fernandes et al., First results of the Phase II SIMPLE dark matter search*, Physical Review Letters* **105** (2010) 211301. doi:10.1103/PhysRevLett.105.211301

[9] T.A. Girard, M. Felizardo, A.C. Fernandes, J.G. Marques and A.R. Ramos, Girard et al. (for the SIMPLE collaboration) reply (to C. E. Dahl et al., preceding Comment, Phys. Rev. Lett. 108, 259001 (2012)), *Physical Review Letters* **108** (2012) 259002. doi:10.1103/PhysRevLett.108.259002

[10] X-5 Monte Carlo Team, MCNP: A General N-Particle Transport Code (Version 5) Vol. 1, LA-UR-03-1987, Los Alamos National Laboratory, 2003.

[11] J. Allison, K. Amako, J. Apostolakis et al., Geant4 developments and applications*, IEEE Transactions on Nuclear Science* **53** (2006) 270-278. doi: 10.1109/TNS.2006.869826

[12] T.T. Böhlen, F. Cerutti, M.P.W. Chin, A. Fassò, A. Ferrari, P.G. Ortega, A. Mairani, P.R. Sala, G. Smirnov and V. Vlachoudis, The FLUKA Code: Developments and challenges for high energy and medical applications, *Nuclear Data Sheets* **120** (2014) 211-214. doi:10.1016/j.nds.2014.07.049.

[13] S. Archambault, F. Aubin, M. Auger et al., Optimization of neutron shielding for the PICASSO experiment, *AIP Conference Proceedings* **1180** (2009) 112-116.

[14] R. Agnese, Z. Ahmed, A. J. Anderson et al., Silicon detector dark matter results from the final exposure of CDMS II, *Physical Review Letters* **111** (2013) 251301. doi:10.1103/PhysRevLett.111.251301.

[15] E. Aprile, M. Alfonsi, K. Arisaka et al., The neutron background of the XENON100 dark matter search experiment, *Journal of Physics G* **40** (2013) 115201. doi: 10.1088/0954-3899/40/11/115201

[16] V.A. Kudryavtsev, M. Robinson and N.J.C. Spooner, The expected background spectrum in NaI dark matter detectors and the DAMA result, *Astroparticle Physics* **33** (2010) 91-96. doi:10.1016/j.astropartphys.2009.12.003.

[17] H.M Araújo, D.Yu. Akimov, E.J. Barnes et al., Radioactivity backgrounds in ZEPLIN–III, *Astroparticle Physics* **35** (2012) 495-502. doi: 10.1016/j.astropartphys.2011.11.001

[18] H.R.T. Wulandari, Study on neutron-induced background in the dark matter experiment CRESST, PhD thesis, Technischen Universität München, 2003.

[19] E. Armengaud, C. Augier, A. Benoit et al., Background studies for the EDELWEISS dark matter experiment, *Astroparticle Physics* **47** (2013) 1-9. doi:10.1016/j.astropartphys.2013.05.004.

[20] C.E. Aalseth, P.S. Barbeau, J. Colaresi et al., CoGeNT: A search for low-mass dark matter using p-type point contact germanium detectors, *Physical Review D* **88** (2013) 012002. doi:10.1103/PhysRevD.88.012002

[21] D.A. Fustin, First dark matter limits from the COUPP 4 kg bubble chamber at a deep underground site, PhD thesis, University of Chicago, 2012.

[22] H. Uchida, K. Abe, K. Hieda et al., Search for inelastic WIMP nucleus scattering on $^{129}$Xe in data from the XMASS-I experiment, *Progress of Theoretical and Experimental Physics* **6** (2014), 063C01. doi: 10.1093/ptep/ptu064

[23] D.S. Akerib, H.M. Araújo, X. Bai et al., Radiogenic and muon-induced backgrounds in the LUX dark matter detector, *Astroparticle Physics* **62** (2015) 33-46. doi:10.1016/j.astropartphys.2014.07.009

[24] H.J. Kim, I.S. Hahn, M.J. Hwang et al., Measurement of the neutron flux in the CPL underground laboratory and simulation studies of neutron shielding for WIMP searches, *Astroparticle Physics* **20** (2004) 549-557. doi: 10.1016/j.astropartphys.2003.09.001

[25] W.B. Wilson, R.T. Perry, W.S. Charlton, T.A. Parish and E.F. Shores, Sources: A code for calculating (alpha,n), spontaneous fission, and delayed neutron sources and spectra, *Radiation Protection Dosimetry* **115** (2005) 117-121. doi: 10.1093/rpd/nci260

[26] M.J. Carson, J.C. Davies, E. Daw et al., Neutron background in large-scale xenon detectors for dark matter searches, *Astroparticle Physics* **21** (2004) 667-687. doi:10.1016/j.astropartphys.2004.05.001

[27] V. Tomasello, V. A. Kudryavtsev and M. Robinson, Calculation of neutron background for underground experiments, *Nuclear Instruments and Methods* **A 595** (2008) 431-438. doi:10.1016/j.nima.2008.07.071

[28] D.-M. Mei, C. Zhang and A. Hime, Evaluation of ($\alpha$,n) induced neutrons as a background for dark matter experiments, *Nuclear Instruments and Methods* **A 606** (2009) 651-660. doi:10.1016/j.nima.2009.04.032. Data available at http://neutronyield.usd.edu

[29] A.C. Fernandes, M. Felizardo, A. Kling, J.G. Marques and T. Morlat, Studies on the efficiency of the neutron shielding for the SIMPLE dark matter search, *Nuclear Instruments and Methods* **A 623** (2010) 960-967. doi:10.1016/j.nima.2010.07.044

[30] A.C. Fernandes, M. Felizardo, T.A. Girard, A. Kling, A.R. Ramos, J.G. Marques, M.I. Prudêncio, R. Marques and F.P. Carvalho, Neutron background estimates in GESA, *E3S Web of Conferences* **4** (2014) 03003. doi: 10.1051/e3sconf/20140403003

[31] M. Felizardo, T.A. Girard, T. Morlat et al., Final analysis and results of the Phase II SIMPLE dark matter search, *Physical Review Letters* **108** (2012) 201302. doi:10.1103/PhysRevLett.108.201302

[32] W.B. Wilson, R.T. Perry, E.F. Shores, et al. SOURCES 4C: A code for calculating (alpha,n), spontaneous fission, and delayed neutron sources and spectra. LA-UR-02-1839, Los Alamos National Laboratory, 2002.

[33] LSBB, http://lsbb.oca.eu

[34] T.A. Girard, F. Giuliani, T. Morlat et al., SIMPLE dark matter search results, *Physics Letters B* **621** (2005) 233-238. doi:10.1016/j.physletb.2005.06.070

[35] F.W. Stallmann, LSL-M2: a computer program for least-squares logarithmic adjustment of neutron spectra, NUREG/CR-4349, ORNL/TM-9933, Oak Ridge National Laboratory, 1985.

[36] Evaluated nuclear data available from the Nuclear Energy Agency at http://www.oecd-nea.org/janisweb/search/endf

[37] L. R. Greenwood and R. K. Smither, SPECTER: Neutron damage calculations for materials irradiations, ANL/FPP/TM-197, Argonne National Laboratory, 1985.

[38] Compositional data from NIST. http://physics.nist.gov/cgi-bin/Star/compos.pl

[39] F.T. Wallenberger, J.C. Watson and H. Li, Glass fibers, in: D.B. Miracle and S.L. Donaldson (Eds.), Composites, ASM Handbook Vol. 21 (American Society for Metals International, Materials Park OH, 2001), pp. 27-34. ISBN 978-0-87170-703-1

[40] B.S. Hayes and L.M. Gammon, Composite materials and optical microscopy, Chapter 1 in: Optical microscopy of fiber-reinforced composites (American Society for Metals International, Materials Park OH, 2010) pp. 1-22. ISBN 978-1-61503-044-6

[41] M. Felizardo, T. Morlat, J.G. Marques, A.R. Ramos, TA Girard, A.C. Fernandes, A. Kling, I. Lázaro, R.C. Martins and J. Puibasset, Fabrication and response of high concentration SIMPLE superheated droplet detectors with different liquids, *Astroparticle Physics* **49** (2013) 28-43. doi: 10.1016/j.astropartphys.2013.08.006

[42] Rousselot gelatin. http://www.parmentier.de/gpfneu/gelatine/rousselot_en.pdf

[43] Duran laboratory glassware. http://www.us.schott.com/labware/english/products/duran/onlinecatalogue/index.html

[44] R.C. Pettersen, The chemical composition of wood, Chapter 2 in: R.M. Rowell (Ed.), The chemistry of solid wood, Advances in Chemistry Series Vol. 27 (American Chemical Society, Washington DC, 1984) pp. 57-126. doi: 10.1021/ba-1984-0207.ch002

[45] E.F. Shores, Data updates for the SOURCES-4A computer code, *Nuclear Instruments and Methods* **A 179** (2001) 78-82. doi:10.1016/S0168-583X(00)00694-7.

[46] F. Manero and V.A. Konshin, Status of the energy-dependent ν-values for the heavy isotopes (Z>90) from thermal to 15 MeV and of ν-values for spontaneous fission, *Atomic Energy Review* **10** (1972) 638-756.

[47] M. Wang, G. Audi, A.H. Wapstra, F.G. Kondev, M. MacCormick, X. Xu and B. Pfeiffer, The AME2012 atomic mass evaluation, *Chinese Physics C* **36** (2012) 1603-2014.

[48] J.F. Ziegler, M.D. Ziegler and J.P. Biersack, SRIM: The stopping and range of ions in matter (2010), *Nuclear Instruments and Methods* **B 268** (2010) 1818-1823. doi:10.1016/j.nimb.2010.02.091

[49] L.S. Waters, G.W. McKinney, J.W. Durkee, M.L. Fensin, J.S. Hendricks, M.R. James, R.C. Johns and D.B. Pelowitz, The MCNPX Monte Carlo radiation transport code, *AIP Conference Proceedings* **896** (2007) 81-90.

[50] A. Zimbal, Measurement of the spectral fluence rate of reference neutron sources with a liquid scintillation detector, *Radiation Protection Dosimetry* **126** (2007) 413-417. doi:10.1093/rpd/ncm085

[51] G.J.H. Jacobs and H. Liskien, Energy spectra of neutrons produced by α-particles in thick targets of light elements, *Annals of Nuclear Energy* **10** (1983) 541-552. doi: 10.1016/0306-4549(83)90003-8

[52] T. Mayer-Kuckuk, Kernphysik, Teubner, Stuttgart, 1984. ISBN 3-519-33021-0

[53] W.N. McElroy, S. Berg, T. Crockett and R. G. Hawkins, A computer-automated iterative method for neutron flux spectra determination by foil activation - Vol. 1, AFWL-TR-67-41, Air Force Weapons Laboratory, 1967.

[54] J.K. Bair and J. Gomez del Campo, Neutron yields for alpha-particle bombardment. *Nuclear Science and Engineering* **71** (1979) 18-28.

[55] D. West and A.C. Sherwood, Measurements of thick-target (α,n) yields from light elements, *Annals of Nuclear Energy* **9** (1982) 551-577. doi:10.1016/0306-4549(82)90001-9

[56] H. Liskien and A. Paulsen, Neutron yields of light elements under α-bombardment. *Atomkernenergie* **30** (1977) 59-61.

[57] G.N. Vlaskin, Yu.S. Khomyakov and V.I. Bulanenko, Neutron yield of the reaction (α, n) on thick targets comprised of light elements, *Atomic Energy* **117** (2015) 357-365. doi:10.1007/s10512-015-9933-5

[58] R. Heaton, H. Lee, P. Skensved and B.C. Robertson, Neutron production from thick-target (α, n) reactions, *Nuclear Instruments and Methods* **A 276** (1989) 529-538. doi:10.1016/0168-9002(89)90579-2

[59] D. West, The calculation of neutron yields in mixtures and compounds from the thick target ($\alpha$,n) yields in the separate constituents, *Nuclear Energy* **6** (1979) 549-552. doi:10.1016/0306-4549(79)90003-3

[60] P. Jagam and J.J. Simpson, Measurements of U, Th and K concentrations in a variety of materials, *Nuclear Instruments and Methods* **A 324** (1993) 389-398. doi: 10.1016/0168-9002(93)91000-D; *Erratum*: NIM A 334 (1993) 657. doi: 10.1016/0168-9002(93)90837-8

[61] J.C. Barton, Studies with a low-background germanium detector in the Holborn underground laboratory, *Nuclear Instruments and Methods* **A 354** (1995) 530-538. doi:10.1016/0168-9002(94)01068-4

[62] Fabrication of borosilicate glass. http://www.madehow.com/Volume-7/Pyrex.html

[63] Boiling points of metals and oxides extracted from http://www.webelements.co.uk

[64] A.C. Fernandes, I. C. Gonçalves, J. Santos, J. Cardoso, L. Santos, A. Ferro Carvalho, J.G. Marques, A. Kling, A.J.G. Ramalho and M. Osvay, Dosimetry at the Portuguese Research Reactor using thermoluminescence measurements and Monte Carlo calculations, *Radiation Protection Dosimetry* **120** (2006) 349-353. doi:10.1093/rpd/nci560

[65] A.C. Fernandes, J.P. Santos, J.G. Marques, A. Kling, A.R. Ramos and N.P. Barradas, Validation of the Monte Carlo model supporting core conversion of the Portuguese Research Reactor (RPI) for neutron fluence rate determinations, *Annals of Nuclear Energy* **37** (2010) 1139-1145. doi: 10.1016/j.anucene.2010.05.004

[66] D.M. Mei and A. Hime, Muon-induced background study for underground laboratories, *Physical Review D* **83** (2006) 053004. doi: 10.1103/PhysRevD.73.053004

[67] Y-F. Wang, V. Balic, G. Gratta, A. Fasso', S. Roesler and A. Ferrari, Predicting neutron production from cosmic-ray muons, *Physical Review D* **64** (2001) 013012. doi:10.1103/PhysRevD.64.013012

[68] F. d'Errico, Radiation dosimetry and spectrometry with superheated emulsions, *Nuclear Instruments and Methods* **B 184** (2001) 229-254. doi: 10.1016/S0168-583X(01)00730-3

[69] M. Barnabé-Heider, M. Di Marco, P. Doane et al., Response of superheated droplet detectors of the PICASSO dark matter search experiment, *Nuclear Instruments and Methods* **A 555** (2005) 184-204. doi: 10.1016/j.nima.2005.09.015

[70] T. Morlat, M. Felizardo, F. Giuliani, et al., A CF3I-based SDD prototype for spin-independent dark matter searches, *Astroparticle Physics* **30** (2008) 159-166. doi:10.1016/j.astropartphys.2008.08.002

[71] T.A. Girard, A.R. Ramos, I.L. Roche and A.C. Fernandes, Phase III (and maybe IV) of the SIMPLE dark matter search experiment at the LSBB. *E3S Web of Conferences* **4** (2014) 01001. doi: 10.1051/e3sconf/20140401001

[72] H.M. Araújo, V.A. Kudryatsev, N.J.C. Spooner and T.J. Sumner, Muon-induced neutron production and detection with GEANT4 and FLUKA, *Nuclear Instruments and Methods A* **545** (2005) 398-411. doi: 10.1016/j.nima.2005.02.004